\documentclass[]{aa}
\usepackage{graphicx}
\begin{document}

\title{X-ray measured metallicities of the intra-cluster medium: a good measure for the metal mass?}

\titlerunning{X-ray measured metallicities of the intra-cluster medium}

\author{W. Kapferer\inst{1}
  \and T. Kronberger\inst{1}
  \and J. Weratschnig\inst{1}
  \and S. Schindler\inst{1}}

\offprints{W. Kapferer, \email{wolfgang.e.kapferer@uibk.ac.at}}

\institute{Institut f\"ur Astro- und Teilchenphysik,
            Universit\"at Innsbruck, Technikerstrasse 25,
            A-6020 Innsbruck}

\date{}

\abstract {}{We investigate whether X-ray observations map heavy
elements in the Intra-Cluster Medium (ICM) well and whether the
X-ray observations yield good estimates for the metal mass, with
respect to predictions on transport mechanisms of heavy elements
from galaxies into the ICM. We further test the accuracy of
simulated metallicity maps.}{We extract synthetic X-ray spectra from
N-body/hydrodynamic simulations including metal enrichment
processes, which we then analyse with the same methods as are
applied to observations. By changing the metal distribution in the
simulated galaxy clusters, we investigate the dependence of the
overall metallicity as a function of the metal distribution. In
addition we investigate the difference of X-ray weighted metal maps
produced by simulations and metal maps extracted from artificial
X-ray spectra, which we calculate with SPEX2.0 and analyse with
XSPEC12.0.}{The overall metallicity depends strongly on the
distribution of metals within the galaxy cluster. The more
inhomogeneously the metals are distributed within the cluster, the
less accurate is the metallicity as a measure for the true metal
mass. The true metal mass is generally underestimated by X-ray
observations. The difference between the X-ray weighted metal maps
and the metal maps from synthetic X-ray spectra is on average less
than 7\% in the temperature regime above $T>3\times 10^{7}$ K, i.e.
X-ray weighted metal maps can be well used for comparison with
observed metal maps. Extracting the metal mass in the central parts
(r$<$500 kpc) of galaxy clusters with X-ray observations results in
metal mass underestimates up to a factor of three.}{}

\keywords{Hydrodynamics -- Methods: numerical -- intergalactic
medium}

\maketitle

\section{Introduction}

Since the first observations of the 7 keV iron line feature in the
1970's by Mitchell et al. (1976) it has been evident that the
intra-cluster medium (ICM) does contain gas already processed by
stars. X-ray spectra are the only measure for the metallicity of the
ICM. With X-ray observatories like XMM-Newton or Chandra it is
nowadays possible to extract metallicities in certain regions of a
galaxy cluster and construct metallicity profiles and X-ray weighted
metallicity maps (Pratt et al. 2006, Durrett et al. 2005). In
addition to observations galaxy cluster simulations including gas
and dark matter (DM) physics are an ideal tool to investigate the
evolution of these systems. By including different enrichment
processes it is possible to trace the metals in the ICM. Recent work
in this field (Cora 2006, Kapferer et al. 2006, Kapferer et al.
2007, Domainko et al. 2006, Moll et al. 2007, Springel \& Hernquist
2003) are able to reproduce the inhomogeneous distribution of metals
within the ICM as observed (Durret et al. 2005). Given the large
mass fraction of the ICM in a cluster (15-20\%) compared to the mass
fraction of the galaxies (3-5\%) a lot of metals must have been
produced within the galaxies and then been transported together with
part of the interstellar medium (ISM) into the ICM. X-ray
observations can also distinguish the lines of different elements,
e.g. elements like Si, S, O from core collapse supernovae or the
elements Fe and Ni from supernovae Ia (Baumgartner et al. 2005,
Ettori et al. 2002, Sanders et al. 2004, Finoguenov et al. 2002) and
can hence give information on the origin of the metals and
interaction in clusters. From these observations we know (e.g.
Mushotzky 1999), that most of the metals are in the ICM and not in
stars. Thus, to trace metal formation and evolution it is crucial to
get correct estimates for the metal content in the ICM.

Three dimensional hydrodynamic galaxy cluster simulations give us an
estimate of the 'true' metallicity distribution within simulated
galaxy clusters. By constructing artificial X-ray profiles and maps
it is possible to test, whether metallicities obtained by X-ray
spectra are a good measure for the real metal content in the ICM or
not.

The paper is structured as follows. First we compare X-ray weighted
metal maps with metal maps obtained from synthetic spectra, which we
construct with SPEX 2.0 (Kaastra et al. 1996) and investigate with
XSPEC 12.0 (Dorman et al. 2003). Then we compare different heavy
element distributions in the ICM and their spectroscopic appearance.
We give constraints on the accuracy of spectroscopic metallicities
as a measure of the real metal content in the ICM. Throughout the
paper we assume solar ratios (Anders \& Grevesse, 1989).

\section{The Model Clusters}

We take two realistic ICM distributions from model clusters for all
further calculations into account. We apply a standard $\Lambda$CDM
simulation with gas physics and constraints random fields (see
Kapferer et al. 2007, Kapferer et al. 2006, Domainko et al. 2006 and
Schindler et al. 2005 for more details) to obtain a reasonable ICM
gas distribution. The first model cluster is a merging galaxy
cluster, hereafter Model Cluster A, whereas the second system is a
more massive galaxy cluster, hereafter Model Cluster B. The
properties of the model galaxy clusters are as follows:

\begin{itemize}

\item \textbf{Model Cluster A:} The cluster forms  at z$\sim$1.5 and has two major
merger events at z=0.8 and z=0.5. The final total mass is
$1.5\times10^{14}$ M$_{\odot}$ within a sphere of radius 1 Mpc.

\item \textbf{Model Cluster B:} The formation redshift for this
cluster is z$\sim$1.7. It shows four minor merger events a z=1.4,
z=1.1, z=0.5 and z=0.3. The cluster has a final mass of
3.4$\times10^{14}$ M$_{\odot}$ within a sphere of radius 1 Mpc.
\end{itemize}

The galaxy cluster simulations have two enrichment processes
included, namely galactic winds and ram-pressure stripping. The
enrichment process description reproduces the metallicity profile of
non-cooling flow and massive galaxy clusters (see Kapferer et al.
(2007) for more details about the enrichment history and a detailed
analysis on the metal distribution). With these two realistic models
for ICM distribution we performed all analysis on the difference of
the true metal mass, which we know exactly from the simulation, and
the metal mass we would obtain from X-ray observations of our model
clusters.

\section{How to construct X-ray weighted metallicity maps}

So far X-ray weighted metal maps where used to obtain metal maps
from simulations. We want to test here how accurate these maps are
compared to maps, that are obtained by using spectra. The total
emissivity by thermal Bremsstrahlung for the ICM above $T>3\times
10^{7}$ K is approximately given by

\begin{equation}
\epsilon^{ff} \sim 1.4 \times 10^{-27} T^{1/2} n_e n_i Z^{2}
g_{B}\;\;\;\; [\rm{ergs\, cm^{-3}\, s^{-1}}],
\end{equation}
\noindent where $T$ is the ICM gas temperature, $n_e$, $n_i$ are the
number densities of electrons and ions, Z is the number of protons
of the bending charge and $g_{B}$ the frequency averaged Gaunt
factor. The emissivity is mainly dominated by the density of the
ICM. Besides the continuum emission several emission lines in the
ICM can be observed. The most prominent lines in the ICM are
typically in the 7 keV iron line complex. These
lines origin from many stages of ionisation of iron.\\
The power emitted by the most prominent emission lines in the ICM,
the iron K line complex, which is a resonance line, is given by

\begin{eqnarray}
P_{jk} = 8.63 \times 10^{6}(n_e n_h A_{el} X_i(t) E_{jk}) \times
\frac{\exp(-E_{gj}/kT)}{T^{1/2}} \nonumber \\  \frac{8
\pi}{\sqrt(3)} f'_{gj} \bar{g}(T)\;\;\;\; [\rm{ergs\, cm^{-3}\,
s^{-1}}]
\end{eqnarray}
\noindent where $A_{el}$ is the abundance of the element $el$
relative to hydrogen and $X_i$ the fraction of the ion $i$, $E_{jk}$
the transition energy from level $j$ to $k$, $E_{gj}$ the transition
energy from ground level $g$ to j, the averaged Gaunt factor
$\bar{g}(T)$ and $f'_{gj}$ the effective oscillator strength. For
more details see Sutherland \& Dopita (1993) and references therein.
Again the dependence of the emission on the density of the ICM and
the temperature of the gas is evident. For a given set of fixed ICM
density and iron abundance the temperature dependence of the
emission ratio of a given iron line and the bremsstrahlung continuum
can be written as

\begin{equation}
R(T)=\frac{P_{ik}}{\epsilon^{ff}}\approx\Omega\frac{\exp(A/T)}{T}
\end{equation}
\noindent where $\Omega$ is an arbitrary normalisation constant and
$A=E_{gj}/k $. Please note that this approximation is different for
each single line of a given ion and that the temperature dependence
of $X_i(T)$, $\bar{g}(T)$ and $g_B(T)$ is ignored here.

\begin{figure}[h]
\begin{center}
{\includegraphics[width=\columnwidth]{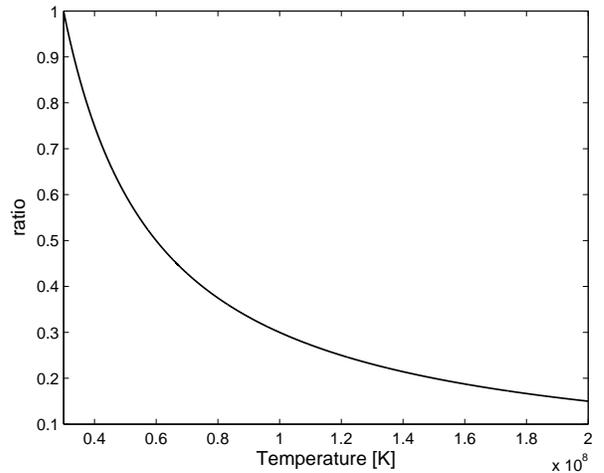}} \caption{The
temperature dependency of Eq. 3, i.e. the ratio of the emissivity
for a given line and the thermal bremsstrahlung continuum in the
temperature range 0.3 - 2$\times 10^{8}$ K.} \label{ratio}
\end{center}
\end{figure}

\noindent In Fig. \ref{ratio} the dependence of a given line and
thermal bremsstrahlung emission as a function of temperature is
given. In our model setup we enrich a fixed space with a fixed mass
of metal. Then we construct X-ray emission weighted metallicities,
see Eq 3. This is done by integrating the metallicities along the
line of sight and the weighting by the square of the density and the
derived temperature dependence, i.e.

\begin{equation}
\rm{Metallicity}_{\rm{X-ray\;weighted}}=\frac{\sum_{los} n^2_{los}
M_{los} R(T)_{los}}{\sum_{los} n^2_{los} R(T)_{los}},
\end{equation}
\noindent where $n_{los}$, $M_{los}$ and $R(T)_{los}$ are the
density, metallicity and temperature-dependent weighting factor in
each cell along the line of sight. By summing also perpendicular to
the line of sight, we can obtain the mean metallicity over the whole
cluster.

\begin{figure}[h]
\begin{center}
{\includegraphics[width=\columnwidth]{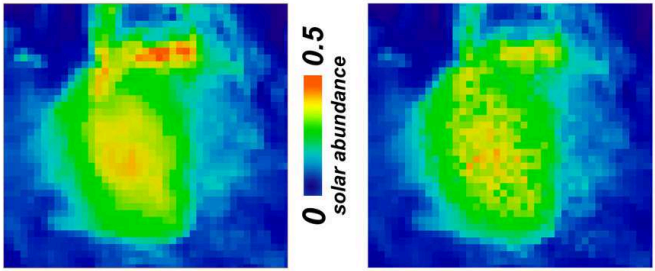}}
\caption{Left: X-ray weighted metal map of the model galaxy cluster
at z=0, the metallicity in each cell is weighted by the X-ray
emission. Right: Synthetic metal map, using a spectrum for each
cell. The image has 1.5 Mpc h$^{-1}$ on the side.}
\label{metal_maps}
\end{center}
\end{figure}

\section{Spectroscopic metal maps and the comparison with X-ray weighted metal maps.}

\begin{figure}[h]
\begin{center}
{\includegraphics[width=\columnwidth]{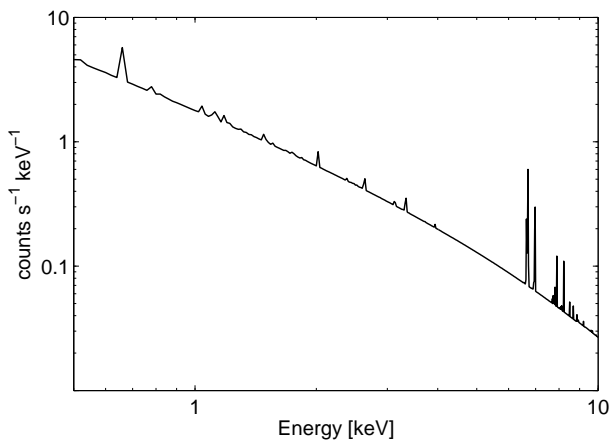}} \caption{Model
spectrum of our model cluster, obtained with SPEX2.0. The count rate
of the whole cluster is shown.} \label{mock_602}
\end{center}
\end{figure}

In order to simulate the X-ray emissivity of our model cluster we
use SPEX 2.0 to construct thermal bremsstrahlung spectra including
line emission. In Fig. \ref{mock_602} a spectrum for the whole model
cluster is given. For each cell of our simulation, which has a
resolution of 128$^{3}$, we calculate a model spectrum which we than
integrate along the line of sight. By using XSPEC we fit a spectrum
to each pixel and hence construct a metallicity map, as shown in
Fig. \ref{metal_maps}.

\begin{figure}[h]
\begin{center}
{\includegraphics[width=\columnwidth]{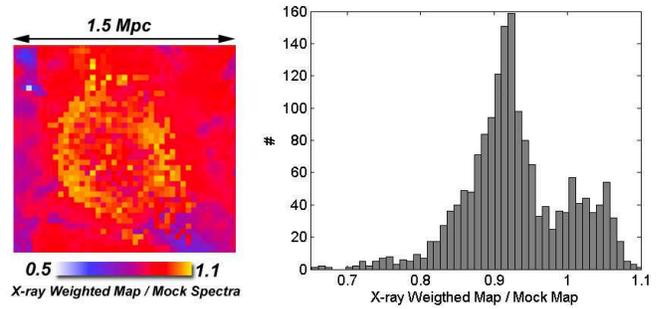}} \caption{Ratio
of X-ray weighted metal map to synthetic spectroscopic metal map.
The image has 1.5 Mpc h$^{-1}$ on the side. Histogram of the left
image. The X-ray weighted metal map differs on average by 7\% from
the spectroscopic metal map.} \label{division}
\end{center}
\end{figure}

The model we used in XSPEC12.0 is exactly the same which we used for
constructing the synthetic spectra. In order to investigate the
quality of our X-ray weighted metal maps we divided the two metal
maps, the metal map obtained with XSPEC12.0 by fitting for each
pixel the synthetic spectra with a MEKAL-model (thermal
bremsstrahlung with line radiation for an optically thin plasma)
(Mewe, Gronenschild \& van den Oord, 1985) and the X-ray weighted
metal map. The result is shown in Fig. \ref{division}, beside the
division image a histogram of the ratios is given. The difference of
the two maps is on average 7\%, leading to the conclusion, that
X-ray weighted metal maps are a good approximation for spectroscopic
metallicity maps, despite all the approximation made. In the
outskirts where the density and temperature of the plasma is lower,
the deviation is larger, due to lower metallicities in the X-ray
weighted metal. This can be understood in terms of the validity of
the thermal bremsstrahlung approximation (Eq. 1), which is only
valid for a plasma with temperatures above $3\times 10^{7}$ K. In
the outskirts of our model cluster r$>$600 kpc the temperature can
drop below $3\times 10^{7}$ K, resulting in a too low metallicity.
In this case the synthetic spectra give better results, because the
Mekal model is valid for plasma temperatures above $3\times 10^{4}$
K (Mewe, Gronenschild \& van den Oord, 1985). In regions 1 Mpc
around massive cluster centres the temperature does typically not
drop below $3\times 10^{7}$ K, therefore the X-ray weighted metal
maps used so far to obtain metal maps from simulated clusters are a
good approximation.

\section{Results}

\subsection{The mean metallicity as a function of the metal
distribution}

\begin{figure}[h]
\begin{center}
{\includegraphics[width=\columnwidth]{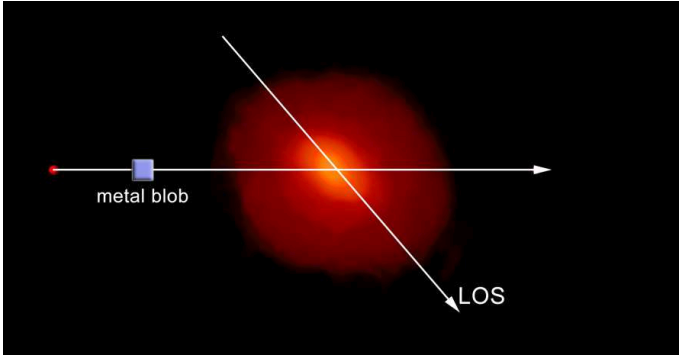}}
\caption{The geometry of the observation and the metal blob
positions. The metal blob is placed at different positions along the
plotted line, which is perpendicular to the line of sight. }
\label{metal1}
\end{center}
\end{figure}

\begin{figure}[h]
\begin{center}
{\includegraphics[width=\columnwidth]{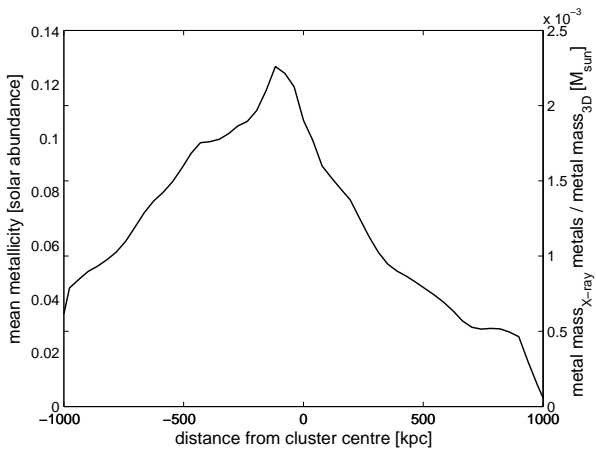}}
\caption{Overall metallicity of the ICM in a circle with 1 Mpc
radius around the cluster centre as a function of the position of a
region containing heavy elements. A metal blob with 117 kpc on a
side containing 1.17x10$^{8}$ M$_{\odot}$ heavy elements moves
through the cluster centre, as highlighted in Fig. \ref{metal1}. On
the right side the ratio of the metal mass obtained from the mean
metallicity and the true 3D metal mass is given. The true metal mass
is always higher then the metal mass obtained from X-ray
observations. Note that the cluster has zero metallicity, expect for
the metal box .} \label{metal2}
\end{center}
\end{figure}

As metallicities in the ICM are always obtained by averaging over
areas or annuli in the field of view, the question arises, how the
real distribution of metals influences the mean metallicity of an
observed cluster. Simulations are the only way to address this
question, because the 3D metal distribution is exactly known. In
order to investigate how sensitive the mean metallicity of the ICM
acts on the distribution of heavy elements we investigate an
extremely inhomogeneous metal distribution in the ICM. We add metals
in a region with $\sim$ 120 kpc on a side containing 1.17x10$^{8}$
M$_{\odot}$ heavy elements, which we place at different positions
along a line perpendicular to the line of sight, which passes
through the cluster centre, i.e. we always add the same amount of
metals to the ICM and see how the mean metallicity of the galaxy
cluster in an 1 Mpc radius area depends on the position of the
metals. Note that the ICM density distribution corresponds to model
cluster A, with zero metallicity, except for the metal box mentioned
above. In Fig. \ref{metal2} the projected mean metallicity of the
ICM in an area with 1 Mpc radius around the cluster centre is shown
as a function of the position of the cube with respect to the
cluster centre. The ratio between the metal mass obtained from the
mean metallicity and the true metal mass is presented as well. The
geometry of the observation and the metal cube position is shown in
Fig. \ref{metal1}. The metallicity measurements always refer to
observations along the line of sight.

It is evident that (see Fig. \ref{metal2}) the position of the cube
influences the result dramatically. This can be understood by the
density distribution within the ICM. The line flux of a given
element can be written as

\begin{equation}
F_{line} \sim \int f(T) n_e n_i dl dA \sim \int f(T) n_e dM_i ,
\end{equation}

where $n_e$, $n_i$ are the number densities of electrons and the
given element ions, $f(T)$ is a temperature dependent emissivity and
$dM_i$ is the differential mass of the given element. In the centre
the high density of the ICM results in the highest mean metallicity,
which contributes most to the X-ray emission. When the metal blob is
placed at a radius of 1 Mpc the mean metallicity drops nearly one
order of magnitude, because the outskirts do contribute less to the
whole cluster emission. From this case of a metal blob placed in
primordial ICM we can conclude, that the mean distribution depends
strongly on the position of the metal blob within the ICM.\\
Recalculating the mass of metals residing in the ICM from the mean
metallicity, shows the same dependence, see Fig. \ref{metal2}. The
mean metallicity always underestimated the metal mass present in the
model cluster. The discrepancy is a factor of 10 in the centre and
increases to a factor of 100 in the outskirts. The reason therefore
is again the fact that emissivity by thermal bremsstrahlung is
mainly driven by the density of the ICM, which decreases from the
cluster centre to the outskirts by several magnitudes and the
inhomogeneous metal distribution. The integration of all quantities
along the line of sight and the investigated area leads to the
discrepancy between the true 3D metal mass and the metal mass
obtained by the mean metallicity of the ICM. This is of course an
extreme example, but it demonstrates how strong the mean metallicity
of a whole cluster depends on the actual metal distribution. In the
case of a constant metallicity over the whole ICM, the metal mass
obtained by the mean metallicity in a circle of 1 Mpc radius around
the cluster centre is of course exactly the same as the true 3D
metal mass.

\subsection{The mean metallicity as a measure for the true metal mass}

\begin{figure}[h]
\begin{center}
{\includegraphics[width=\columnwidth]{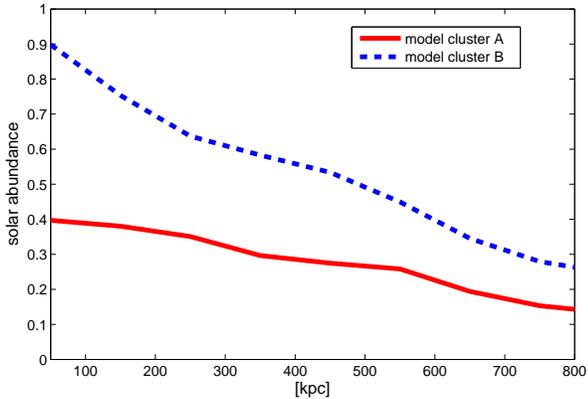}}
\caption{Metallicity profiles for model clusters A and B.}
\label{metal_profiles}
\end{center}
\end{figure}

\begin{figure*}[h]
\begin{center}
{\includegraphics[width=\textwidth]{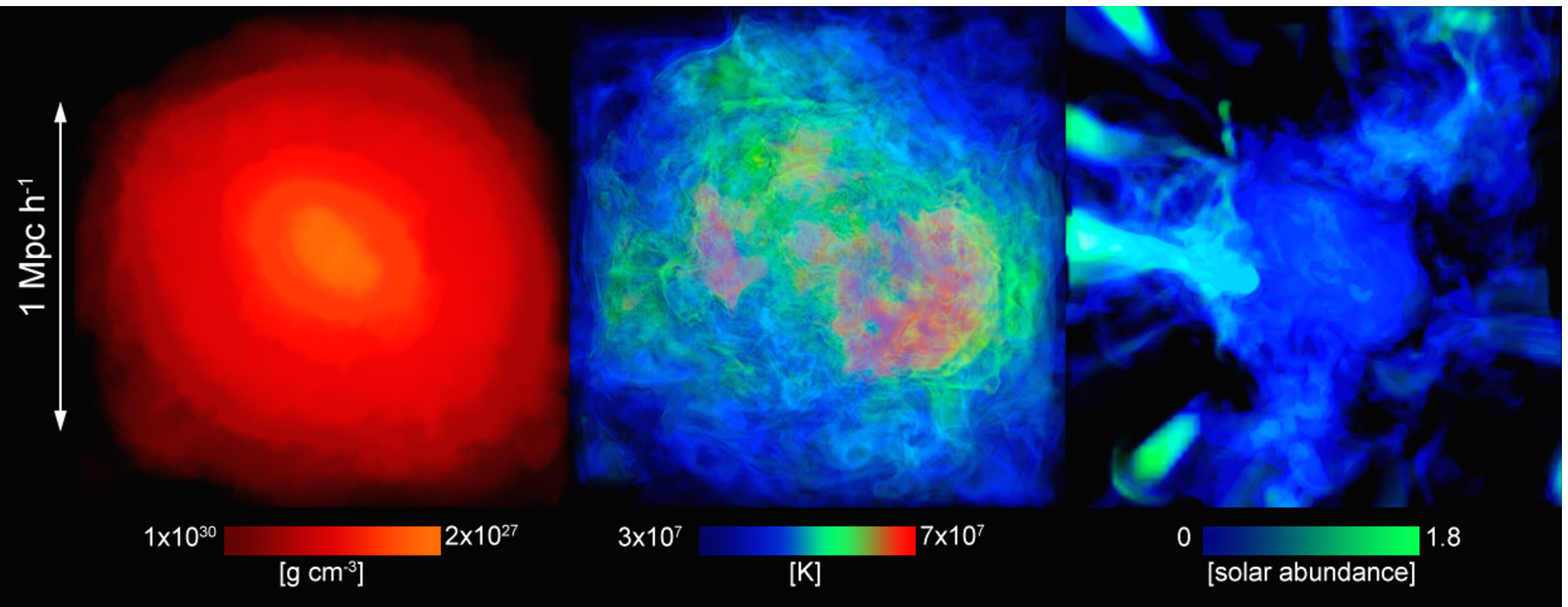}} \caption{Three
dimensional distribution of the density, temperature and the
metallicity of the ICM of model cluster A.} \label{threeD}
\end{center}
\end{figure*}

In order to test, whether the mean metallicity in a defined region
around the cluster centre represents a good measure for the true
metal mass within this area or not, we performed several tests with
the model clusters. The metal content in the model galaxy clusters
were calculated from an enrichment process investigation, where
galactic winds and ram-pressure stripping acts simultaneously to
enrich the ICM. The simulation starts at a redshift of $\sim$20 and
results in a metal distribution within the ICM as shown in Fig.
\ref{metal_profiles}. For more details on the enrichment processes
and the metal distribution see Kapferer et al. (2007). At redshift
$z=0$ we extract the mean metallicity within circles of different
radii around the cluster centre (100, 200, 500 and 800 kpc) and
extract from the 3D simulation the mean ICM density within the
corresponding spheres. As we know exactly the amount and
distribution of heavy elements, we can compare these two quantities,
i.e. the true heavy element mass and the X-ray measured heavy
element mass. In Fig. \ref{threeD} the spatial distribution of the
density, temperature and metallicity of model cluster A is
presented. Whereas the density does not show significant
substructures, the metallicity is inhomogeneous.

\begin{figure}[h]
\begin{center}
{\includegraphics[width=\columnwidth]{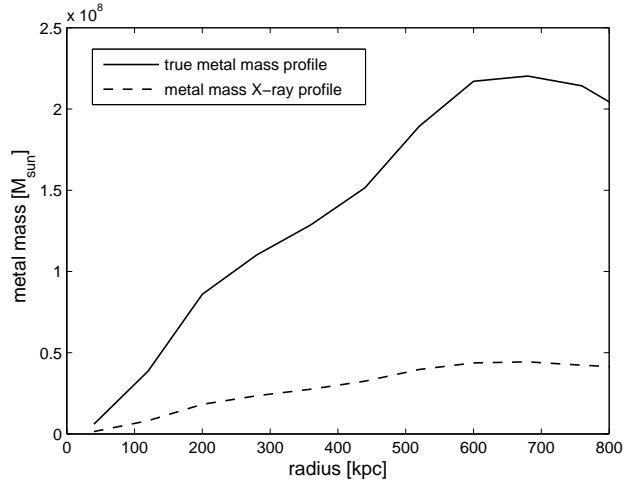}}
\caption{Metal mass profile per radial bin for model cluster A.
Solid line: true metal mass within a three dimensional radial bin.
Dashed line: metal mass obtained by multiplying the metallicity
profile taken from the X-ray weighted metal map (eqn. 4) with the
three dimensional gas mass within the same bin (bin size 80 kpc).}
\label{metal_mass_profile}
\end{center}
\end{figure}

In Table \ref{metals_table} the mean metallicity, true metal and
X-ray measured metal mass for different circles around the cluster
centre for model cluster A and B are given. We find, that the X-ray
measured metal mass always results in lower heavy element masses
compared to the true metal mass in the same region of interest. If
the radius is smaller, the discrepancy can be up to three times more
metals in the ICM, than measured by X-ray spectra. By increasing the
radius, i.e. averaging over much larger volumes, the discrepancy
becomes smaller. At a radius at 800 kpc the X-ray measured heavy
element mass gives 20\% to 50\% lower masses than the true metal
mass. In Fig. \ref{metal_mass_profile} two metal-mass profiles for
model cluster A are shown. The metal mass increases in general with
increasing radius, because the volume of the radial shell increases
with the square of the distance to the cluster centre. Although the
metallicity decreases at larger radii, the larger volume results in
a higher metal mass. The metal mass X-ray profile (dashed line) is
considerably lower then the true metal mass profile. The discrepancy
arises by averaging of the metallicity in annuli at all radii, which
wipes out very prominent metallicity peaks all over the cluster. In
Fig. \ref{threeD} the distribution of the density, temperature and
metallicity of the ICM in model cluster A at redshift 0 is shown.
The high metallicity regions in the outskirts are clearly visible in
the 3D metal distribution. The inhomogeneous metal distribution in
galaxy clusters leads to a systematic underestimate of the metal
mass by X-ray observations. For example the ratio of the integrated
true metal-mass profile to the integrated X-ray metallicity obtained
metal-mass profile in an 200 kpc radius is 2.70. Investigations of
the amount of metals residing in galaxies in galaxy clusters and in
the ICM (e.g. Renzini et al. 1993) do always underestimate the true
metal mass within the ICM. These works concluded, that on average
the amount of heavy elements within galaxies and the ICM is
comparable, would get a shift towards the conclusion, that even more
metal mass resides within the ICM as in the galaxies.

\begin{table*}
\begin{center}
\caption[]{True metal mass versus X-ray obtained metal mass in
different spheres around the cluster centre for a less massive model
galaxy cluster A and a massive model galaxy cluster B.}
\begin{tabular}{ c c c c c c }
\hline \hline  & & & & Mean metallicity & Ratio \cr Model cluster &
Radius [kpc] & True metal mass [M$_{\odot}$] & X-ray measured metal
mass [M$_{\odot}$] & Solar abundance & C.3 to C.4\cr \hline A & 800
& $1.07\times10^{10}$ & $8.40\times10^{9}$ & 0.346 & 1.27 \cr A &
500 & $4.50\times10^{9}$ & $2.63\times10^{9}$ & 0.356 & 1.71 \cr A &
200 & $6.05\times10^{8}$ & $2.24\times10^{8}$ & 0.370 & 2.70 \cr A &
100 & $9.70\times10^{7}$ & $3.10\times10^{7}$ & 0.383 & 3.13\cr
\hline B & 800 & $1.02\times10^{11}$ & $6.70\times10^{10}$ & 0.94 &
1.52 \cr B & 500 & $4.70\times10^{10}$ & $2.75\times10^{10}$ & 0.95
& 2.80\cr B & 200 & $7.50\times10^{9}$ & $3.45\times10^{9}$ & 0.96 &
2.17 \cr B & 100 & $1.25\times10^{9}$ & $5.40\times10^{8}$ & 0.97 &
2.31\cr \hline \hline
\end{tabular}
\label{metals_table}
\end{center}
\end{table*}

\section{Discussion and Conclusions}

We investigate the quality of the metallicities obtained by X-ray
observation of the ICM as a measure for the true metal mass of the
ICM. In addition we test if X-ray weighted metal maps of galaxy
cluster simulations are comparable to synthetic X-ray spectra
composed of thermal bremsstrahlung and line radiation. The results
of our work are as follows:

\begin{itemize}

\item If the ICM temperature is above $T>3\times 10^{7}$ K X-ray
weighted metal maps results nearly in the same metallicities as
obtained from synthetic X-ray spectra. The difference is on average
less than 7\% in areas with $r<750$ kpc around the centre of galaxy
clusters.

\item We test how much the distribution of metals within a
sphere of r$=1$ Mpc around the cluster centre changes the X-ray
measured mean metallicity of this area. We found that the X-ray
obtained mean metallicity can change by a factor of 10 in the
extreme case of a distinct metal blob residing in primordial ICM,
depending on the distance from the cluster centre.

\item By using realistic metal distributions in simulated galaxy
clusters we test the quality of the X-ray obtained overall
metallicity as a measure for the metal mass. We found that depending
on the size of the investigated area around the cluster centre the
true metal mass can be three times higher than the metal mass
obtained by X-ray observations. If the radius increases the
discrepancy gets smaller, but in all cases the true metal mass is
higher than the metal mass obtained by X-ray observations.

\end{itemize}

The discrepancies of the true metal mass and the X-ray obtained
metal mass within the ICM in galaxy clusters can be understood in
terms of averaging. The observer measures integrated quantities
along the line of sight, including dense regions in the centre and
less dense regions in the outskirts. In order to be able to get
reasonable X-ray spectra of the ICM, observers must always collect
X-ray photons over an area. As the metallicity is not constant
throughout the galaxy cluster, i.e. the ratio of primordial ICM to
heavy elements is not constant, the integration of thermal
bremsstrahlung and line radiation results in different ratios,
leading to too low metal masses. Applying multi-temperature instead
of single-temperature models for the ICM would alleviate the
problem partially as shown by Buote \& Canizares (1994).\\
Taking the results of this work into account, it is obvious that
estimates from X-ray observations on the heavy elements (e.g. iron)
content in the ICM lead to masses, which are too low.

\section*{Acknowledgements}

The authors thank the anonymous referee for fruitful comments, which
helped to improve the paper. The authors would like to thank Etienne
Pointecouteau for the useful help with XSPEC12.0 and Jelle Kaastra
for the help regarding SPEX2.0. The authors acknowledge the Austrian
Science Foundation (FWF) through grants P18523-N16 and P19300-N16.
Thomas Kronberger is a recipient of a DOC-fellowship of the Austrian
Academy of Sciences. In addition, the authors acknowledge the ESO
Mobilit\"atsstipendien des BMWF (Austria), the Tiroler
Wissenschaftsfonds (Gef\"ordert aus Mitteln des vom Land Tirol
eingerichteten Wissenschaftsfonds) and the UniInfrastrukturprogramm
2005/06 from the BMWF.

\end{document}